\documentclass[
reprint,
 amsmath,amssymb,
 aps,
]{revtex4-2}
\usepackage{overpic}
\usepackage{makecell}
\usepackage{graphicx}
\usepackage{epsfig}
\usepackage{dcolumn}
\usepackage{bm}
\usepackage{rotating}
\usepackage{multirow}
\usepackage{subfigure}
\usepackage{hhline}
\usepackage{amssymb}
\usepackage{xcolor}

\usepackage{lineno}
\usepackage{hyperref}
\hypersetup{hypertex=true,colorlinks,linkcolor=blue,anchorcolor=blue,citecolor=blue,urlcolor=blue}
\usepackage{float}
\usepackage{hyperref}
\newcommand\Eone{E_{\rm \uppercase\expandafter{\romannumeral1}}}
\newcommand\Etwo{E_{\rm \uppercase\expandafter{\romannumeral2}}}

\newcommand\bes{BES\uppercase\expandafter{\romannumeral3}}
\newcommand\BEPC{BEPC\uppercase\expandafter{\romannumeral2}}

\let\oldequation\equation
\let\oldendequation\endequation
\renewenvironment{equation}
  {\linenomathNonumbers\oldequation}
  {\oldendequation\endlinenomath}

\let\oldgather\gather
\let\oldendgather\endgather
\renewenvironment{gather}
  {\linenomathNonumbers\oldgather}
  {\oldendgather\endlinenomath}

\usepackage[normalem]{ulem}

\begin{document}

\preprint{BESIII/XYZ}

\title{Search for $\eta_c(2S)\to p\bar{p}K^+K^-$ and measurement of $\chi_{cJ}\to p\bar{p}K^+K^-$ in $\psi(3686)$ radiative decays}

\author{
M.~Ablikim$^{1}$, M.~N.~Achasov$^{4,c}$, P.~Adlarson$^{76}$, O.~Afedulidis$^{3}$, X.~C.~Ai$^{81}$, R.~Aliberti$^{35}$, A.~Amoroso$^{75A,75C}$, Y.~Bai$^{57}$, O.~Bakina$^{36}$, I.~Balossino$^{29A}$
% ...
\\
\vspace{0.2cm}
(BESIII Collaboration)\\
\vspace{0.2cm} {\it
$^{1}$ Institute of High Energy Physics, Beijing 100049, People's Republic of China\\
% ...
$^{3}$ Bochum  Ruhr-University, D-44780 Bochum, Germany\\
$^{4}$ Budker Institute of Nuclear Physics SB RAS (BINP), Novosibirsk 630090, Russia\\
% ...
$^{29}$ INFN Sezione di Ferrara, (A)INFN Sezione di Ferrara, I-44122, Ferrara, Italy; (B)University of Ferrara,  I-44122, Ferrara, Italy\\
% ...
$^{35}$ Johannes Gutenberg University of Mainz, Johann-Joachim-Becher-Weg 45, D-55099 Mainz, Germany\\
$^{36}$ Joint Institute for Nuclear Research, 141980 Dubna, Moscow region, Russia\\
% ...
$^{57}$ Southeast University, Nanjing 211100, People's Republic of China\\
% ...
$^{75}$ University of Turin and INFN, (A)University of Turin, I-10125, Turin, Italy; (B)University of Eastern Piedmont, I-15121, Alessandria, Italy; (C)INFN, I-10125, Turin, Italy\\
$^{76}$ Uppsala University, Box 516, SE-75120 Uppsala, Sweden\\
% ...
$^{81}$ Zhengzhou University, Zhengzhou 450001, People's Republic of China\\
\vspace{0.2cm}
% ...
$^{c}$ Also at the Novosibirsk State University, Novosibirsk, 630090, Russia\\
% ...
}
}

\begin{abstract}
  A search for $\eta_c(2S)\to p\bar{p}K^+K^-$, together with
  measurement of branching fractions of
  $\chi_{cJ(J=0,1,2)}\to p\bar{p}K^+K^-$ in the
  $\psi(3686) \to \gamma \eta_c(2S)$ and the
  $\psi(3686) \to \gamma \chi_{cJ}$ radiative decays, is performed
  with $(2712.4\pm14.3)\times 10^6$ $\psi(3686)$ events collected with
  the \bes~detector at the BEPC\uppercase\expandafter{\romannumeral2}
  collider. An evidence for $\eta_c(2S)\to p\bar{p}K^+K^-$ is found,
  with a significance of $3.3\sigma$. The product
  branching fraction of
  $\mathcal{B}[\psi(3686)\to\gamma\eta_c(2S)]\cdot\mathcal{B}[\eta_c(2S)\to
  p\bar{p}K^+K^-]$ is determined to be
  $(1.98\mkern 2mu\pm\mkern 2mu0.41_{\text{stat.}}\mkern 2mu\pm\mkern
  2mu0.99_{\text{syst.}})\times 10^{-7}$.  The product branching fractions of
  $\mathcal{B}[\psi(3686)\to\gamma\chi_{cJ}]\cdot\mathcal{B}[\chi_{cJ}\to
  p\bar{p}K^+K^-]$ are measured to be
  $(2.49\mkern 2mu\pm\mkern 2mu 0.03_{\text{stat.}}\mkern 2mu\pm\mkern
  2mu 0.15_{\text{syst.}})\times 10^{-5}$,
  $(1.83\mkern 2mu \pm\mkern 2mu 0.02_{\text{stat.}}\mkern 2mu
  \pm\mkern 2mu 0.11_{\text{syst.}})\times 10^{-5}$, and
  $(2.43\mkern 2mu\pm\mkern 2mu 0.02_{\text{stat.}}\mkern 2mu\pm\mkern
  2mu 0.15_{\text{syst.}})\times 10^{-5}$, for $J=0,\ 1$, and 2,
  respectively.
\end{abstract}
\maketitle

\section{Introduction}
The charmonia below the open-charm threshold are well estabilished,
and their spectrum can be well described by the Quantum Chromodynamics
(QCD) or QCD-inspired models. The study of these charmonia would offer
valuable insights into QCD, and provide valuable reference for
decoding the nature of many exotic candidates above the open-charm
threshold.  However, the decay dynamics of the charmonia is far from
being well understood.

The ``$\rho\pi$ puzzle" remains an outstanding issue for the vector
charmonia~\cite{Mo:2006cy}. The puzzle stems from the theoretical
prediction that the ratio $Q_V$, defined by
\begin{equation}
Q_V=\frac{\mathcal{B}[\psi(3686)\to \text{hadrons}]}{\mathcal{B}[J/\psi\to \text{hadrons}]},\label{eq_qv}
\end{equation}
should be about $12\%$~\cite{Appelquist:1974zd}.This prediction is
borne out by many measurements of exclusive decay modes, but there are
some anomalous results from the $\rho\pi$ and other decay
modes~\cite{Franklin:1983ve}. A similar ratio is also proposed for
pseudoscalar charmonia, denoted as $Q_P$,
\begin{equation}
Q_P=\frac{\mathcal{B}[\eta_c(2S)\to \text{hadrons}]}{\mathcal{B}[\eta_c\to \text{hadrons}]}.\label{eq_qp}
\end{equation}
Yet, the available expectation values for $Q_P$ from different
theories are inconsistent~\cite{Anselmino:1991es,Chao:1996sf}.

By analyzing existing experimental data on pseudoscalar charmonium
decays, a combined fit~\cite{Wang:2021dxw} is performed to retrieve
$Q_P$. The result indicates that the fitted ratios from the
experimental data are significantly different from both of the
theoretical models~\cite{Anselmino:1991es,Chao:1996sf}, and the
branching fractions of almost all exclusive decay modes of
$\eta_{c}(2S)$ are smaller than those of $\eta_c$. These discrepancies
could hint at the intriguing decay dynamics of $\eta_c$ and
$\eta_c(2S)$.

Compared to $J/\psi$ and $\psi(3686)$, the states of $\eta_c$ and
$\eta_c(2S)$ are relatively less studied, especially for the
latter. The $\eta_c(2S)$ was first observed by Belle in $B$
decays~\cite{Belle:2002bnx} in 2002, and was subsequently corroborated
by BaBar and CLEO~\cite{BaBar:2003osb,CLEO:2003gwz,BaBar:2011gos} in
two-photon fusion reactions. The $\eta_c(2S)$ can also be produced in
the magnetic dipole (M1) transition of $\psi(3686)$. Due to the small
branching fraction of the M1
transition~\cite{Barnes:2005pb,Li:2007xr,Peng:2012tr,Li:2011ssa,Negash:2017rqt},
a large $\psi(3686)$ sample is needed. Moreover, good performance of
the electromagnetic calorimeter (EMC) is required to detect the
low-energy M1 photon. It was not until 2012 that the $\eta_c(2S)$ from
M1 transition of $\psi(3686)$ was firstly observed by BESIII in
$\eta_c(2S)\to K^0_sK^{\pm}\pi^{\mp}$ and $\eta_c(2S)\to K^+K^-\pi^0$
decay modes~\cite{BES:2012uhz}, with a statistical significance
greater than $10\sigma$. So far, as compiled by the Particle Data
Group (PDG), the measured branching fractions of $\eta_c(2S)$ only add
up to roughly $7\%$~\cite{ParticleDataGroup:2024cfk}. Further investigations
into its additional decay modes are needed.

In this paper, we perform a search for the
$\eta_c(2S)\to p\bar{p}K^+K^-$ process in the radiative decay
$\psi(3686)\to\gamma\eta_c(2S)$ with $(2712.4\pm14.3)\times 10^6$
$\psi(3686)$ events collected by BESIII in 2009, 2012, and
2021~\cite{BESIII:2024lks}. The $\chi_{cJ(J=0,1,2)}$ states, the
$P$-wave spin-triplet charmonia, have been observed to decay into the
same final state with 106 million $\psi(3686)$
events~\cite{Wang:2011kti}. The $\chi_{cJ}\to p\bar{p}K^+K^-$ are
dominant backgrounds for $\eta_c(2S)\to p\bar{p}K^+K^-$, so improved measurements
of these decays are important to extract a
reliable yield of the $\eta_c(2S)$ decay. The product branching
fractions of $\psi(3686)\to\gamma\chi_{cJ}\to\gamma p\bar{p}K^+K^-$
are measured for the first time in this paper.

\section{Besiii detector and Monte Carlo Simulation}
The BESIII detector at BEPCII is designed to study hadron spectroscopy
and $\tau$-charm physics~\cite{BESIII:2009fln}. The cylindrical
BES\uppercase\expandafter{\romannumeral3} is composed of a Helium-gas
based multilayer drift chamber (MDC) with superconducting quadrupoles
inserted in the conical shaped end caps, a Time-of-Flight (TOF) system
located outside the MDC, a CsI(Tl) EMC placed outside of the TOF
system and inside the superconducting solenoid, providing 1.0 T
magnetic field in the central region of
BES\uppercase\expandafter{\romannumeral3}, a muon identifier outside
the superconducting solenoid consisting of layers of resistive plate
chambers inserted in gaps between steel plates of the flux return
yoke. The momentum resolution of charged particles at 1 GeV is $0.5\%$
and the d$E$/d$x$ resolution is $6\%$ for the electrons from Bhabha
scattering at 1~GeV. The photon energies are measured by the EMC with
a resolution of $2.5\%$ ($5\%$) at 1 GeV in the barrel (end-cap)
region. The time resolution of the barrel (end-cap) TOF system is 68
ps (110 ps). The end-cap TOF system was upgraded in 2015 using the
multi-gap resistive plate chamber technology, improving the time
resolution from 110 ps to 60
ps~\cite{Li:2017jpg,Guo:2017sjt,Cao:2020ibk}, which benefits $83\%$ of
the data used in this analysis.

Simulated Monte Carlo (MC) samples are produced by
\textsc{geant4}-based simulation software
\textsc{boost}~\cite{GEANT4:2002zbu,ZYDeng:2006}, which includes the
geometric description of the BESIII detector, as well as the running
conditions and response. The production of $\psi(3686)$ is simulated
using \textsc{kkmc}~\cite{ref:kkmc1,ref:kkmc2}, and the hadronic decay
of $\psi(3686)$ is generated by \textsc{evtgen}~\cite{evtgen1,evtgen2}
with branching fractions of known decay modes from PDG, or by
\textsc{lundcharm} for unobserved decay modes. The final state
radiation (FSR) from charged particles is incorporated with the
\textsc{photos} package~\cite{photos}.

To study the detection efficiencies, exclusive MC samples of
$\psi(3686)\to\gamma\chi_{cJ}$, with $\chi_{cJ}$ decays to
$p\bar{p}K^+K^-$ and $\chi_{cJ}\to I\to p\bar{p}K^+K^-$ are generated
by the phase-space model, where $I$ stands for intermediate states
including $\Lambda(1520)\bar{\Lambda}(1520)$,
$\Lambda(1520)\bar{p}K^++c.c.$, and $p\bar{p}\phi$. The efficiencies
of $\psi(3686)\to\gamma\chi_{cJ}\to\gamma p\bar{p}K^+K^-$ are obtained
from a synthesis MC sample in which these processes with different
decay dynamics are mixed according to the branching fractions measured
previously with the 106 million $\psi(3686)$
events~\cite{Wang:2011kti}. An exclusive MC sample of
$\psi(3686)\to\gamma\eta_c(2S)\to\gamma p\bar{p}K^+K^-$, where the
$\eta_c(2S)$ decay is generated with the phase-space model, is used to
estimate the efficiency of the $\eta_c(2S)$ decay signal. In the
production processes of $\psi(3686)\to\gamma\chi_{cJ}$ and
$\gamma\eta_c(2S)$, the photon follows the angular distribution of
$1+\alpha\cos^2{\theta_\gamma}$, where $\theta_\gamma$ is the polar angle of the
photon's momentum in the rest frame of $e^+e^-$. From the conservation of C and P parities, $\alpha$ is 1 for $\eta_c(2S)$. Assuming $\chi_{c0}$ and $\chi_{c1}$ are produced in electric dipole (E1) transition, $\alpha$ is 1 and $-1/3$ for $\chi_{c0}$ and $\chi_{c1}$, respectively. Experimental measurements from Ref.~\cite{BES:2004ijm} are used for $\chi_{c2}$, from which the $\alpha$ is about $1/12$. An exclusive
background MC sample of $\psi(3686)\to p\bar{p}K^+K^-$ and an
inclusive MC sample comprising 2.747 billion $\psi(3686)$ events are
utilized for background study. The inclusive MC sample includes the
production of the $\psi(3686)$ resonance, the initial-state radiation
production of the $J/\psi$ meson, and the continuum processes
incorporated in \textsc{kkmc}. The \textsc{topoana}
package~\cite{Zhou:2020ksj} is used to analyze the decay chains of
events in the inclusive MC sample.

\section{Event selection and background analysis}
Candidate events must have four charged tracks and at least one
photon. Charged tracks in the MDC are required to satisfy
$\vert\!\cos\theta\vert<0.93$, where $\theta$ represents the polar angle with
respect to the positron beam direction. Events with two oppositely
charged track pairs are selected. The minimum distance of the
closest points of the tracks to the interaction point is required to be less
than 1 cm in the transverse plane and less than 10 cm in the beam
direction. Particle identification (PID) is performed on each charged
track, combining the TOF and d$E$/d$x$ information. The PID hypothesis
with the highest probability is assigned to the track, where we
require a minimum PID confidence level (C.L.) of $0.1\%$. After the
PID, we select events containing exactly one $p$, $\bar{p}$, $K^+$,
and $K^-$.

For identification of photon candidates, the deposited energy of each
shower in the EMC must be larger than 25~MeV in the barrel region
($\vert\!\cos\theta\vert<0.80$), and larger than 40 MeV in the end-cap region
($0.86<\vert\!\cos\theta\vert<0.92$). The interval between the EMC time and
the event start time is required to be within $[0, 700]$ ns for
noise suppression.  A vertex fit is performed with the four charged
tracks, and only events with successful vertex fits are retained. Then
a kinematic fit with four constraints (4C) is performed to ensure
energy-momentum conservation between the initial and final states. If
there are multiple photon candidates, the one with the lowest
$\chi^2_{\rm 4C}$ value is selected as the photon emitted from
$\psi(3686)$. A subsequent three-constraint (3C) kinematic fit is
carried out, where the energy of the photon is free to vary. This 3C
kinematic fit can provide a better separation of $\eta_c(2S)$ signal
from the background.

The $\chi^2_{\mathrm{4C}}$ criterium is optimized by maximizing the figure of merit (FOM) given by
\begin{equation}
  \text{FOM}_a = \frac{S}{\frac{b^2}{2}+a\sqrt{B}+\frac{b}{2}\sqrt{b^2+4a\sqrt{B}+4B}},\label{eq:fom1} \end{equation}
following the discussion in Ref.~\cite{Punzi:2003bu}.  In
Eq.~(\ref{eq:fom1}), $B$ is the number of background events in the
$\chi_{c2}$ mass region ([3.53, 3.58] $\text{GeV}/c^2$), obtained from
the inclusive MC sample. The symbol $S$ represents the number of
signal events in the same region, obtained using the exclusive MC
sample of $\psi(3686)\to\gamma\chi_{c2}\to\gamma p\bar{p}K^+K^-$,
normalized according to the branching fractions of $\chi_{c2}\to
p\bar{p}K^+K^-$, $\chi_{c2}\to \Lambda(1520)\bar{p}K^++c.c.$,
$\chi_{c2}\to \Lambda(1520)\bar{\Lambda}(1520)$, and $\chi_{c2}\to
p\bar{p}\phi$ from the PDG. The parameter $a$ represents the significance level, characterizing
the probability of rejecting the null hypothesis when it is true, and
$b$ describes the confidence level, such that the value of $1-{\rm C.L.}$ is the probability of rejecting the alternative hypothesis. The optimization is stable for
different combinations of $a$ and $b$.

There is a discrepancy between the MC and data samples 
for the fake photons
which denote photons not selected by the 4C kinematic fit. This
discrepancy manifests in the distribution of
$\theta(\gamma, \bar{p})-E_{\gamma}$, where $E_{\gamma}$ is the energy
of the fake photon and $\theta(\gamma, \bar{p})$ is the angle between
the three-momentum of the fake photon and the $\bar{p}$. This discrepancy
will affect the efficiency evaluated by the MC simulation. Using
$\psi(3686)\to\gamma\chi_{c2}\to\gamma p\bar{p}K^+K^-$ as control
sample, we compare the two dimensional distribution between the
control sample and the corresponding MC sample. The
$\theta(\gamma, \bar{p})$ is divided into bins with $10^{\circ}$ width
and $E_{\gamma}$ is divided into bins with 25 MeV width. The ratio
$r=n^{\rm Data}/n^{\rm MC}$ is calculated for each bin, where
$n^{\rm Data(MC)}$ is the number of fake photons in data (normalized
MC sample) in one bin. The truth matching is performed to the photons
before the 4C kinematic fit in all MC samples. Fake photons identified in
the MC sample are re-sampled according to the ratio $r$ to match the yields
in data as a function of $E_{\gamma}$ and $\theta(\gamma,
\bar{p})$. The re-sampled photon candidates in the MC samples are then
subject to the 4C kinematic fit. After this procedure, the
$\theta(\gamma, \bar{p})-E_{\gamma}$ distribution of fake photons in
MC sample is consistent with that in the data, while other distributions like
$\theta(\gamma, p)$, $\theta(\gamma, K^{\pm})$ are not affected.

The study of the inclusive MC sample indicates the presence of background
from the $\psi(3686)\to p\bar{p}K^+K^-$ process, where an FSR photon
or a spurious photon from the detector is misidentified as the one
directly from the $\psi(3686)$ radiation. In the invariant mass
spectrum of $p\bar{p}K^+K^-$ from the 3C kinematic fit, this background
peaks at the mass of $\psi(3686)$ and extends to
3.6~GeV. The lineshape of $\psi(3686)\to p\bar{p}K^+K^-$ depends on
the fraction of FSR process $f_{\rm FSR}$, defined as
\begin{equation}
f_{\rm FSR}=\frac{N_{\rm FSR}}{N_{\rm nonFSR}},
\end{equation}
where the numerator and the denominator are the numbers of FSR and
non-FSR event yields, respectively. The fraction $f_{\rm FSR}$ is
usually different for the MC and data. Using the control sample of
$\psi(3686)\to\gamma\chi_{c0}\to\gamma \gamma_{\rm
  FSR}p\bar{p}K^+K^-$, the ratio
$R_{\rm FSR}=f_{\rm FSR}^{\rm Data}/f_{\rm FSR}^{\rm MC}=2.38\ \pm\
0.90$ is determined, where the uncertainty is statistical only. The
fraction of FSR events in the exclusive $\psi(3686)\to p\bar{p}K^+K^-$
process is reweighted by $R_{\rm FSR}$. The data set at $\sqrt{s}=3.65$
GeV with integrated luminosity $401\ \rm{pb}^{-1}$ is utilized to
study potential backgrounds from the continuum production, which is
found to be negligible.

\section{Determination of branching fractions}
To obtain the signal yields, an unbinned maximum likelihood fit is
performed to the invariant mass of $p\bar{p}K^+K^-$ from the 3C
kinematic fit ($M_{p\bar{p}K^+K^-}^{\rm 3C}$). The full probability
density function (PDF) is the incoherent sum of three components for
the $\eta_c(2S)$ signal, the $\chi_{cJ}$ signal, and the
$\psi(3686)\to p\bar{p}K^+K^-$ background. The PDF of the $\eta_c(2S)$
signal is described as
\begin{equation}
    \frac{{\rm d}\Gamma}{{\rm d}M}\epsilon(M)\otimes F_{\rm res}^{(a)}(\delta m_1,\sigma_1)\otimes F_{\rm res}^{(b)}(\delta m_2,\sigma_2),
\label{eq6}
\end{equation}
Here, $\epsilon$ is the energy-dependent efficiency obtained by
applying the Gaussian Process Regression
(GPR)~\cite{RasmussenWilliams:2006gpml} to the discrete distribution
of the efficiency extracted from the exclusive MC sample of the
$\eta_c(2S)$ signal. The differential decay width is given by
\begin{equation}
  \label{eq7}
    \frac{{\rm d}\Gamma}{{\rm d}M}=\frac{2M^2}{\pi}\frac{E_{\gamma}^3\mathcal{F}(E_{\gamma}) \Gamma^{\text{decay}}(M)}{(M^2-m_{\eta_{c}(2S)}^2)^2+m_{\eta_{c}(2S)}^2\Gamma_{\eta_c(2S)}^2}.
\end{equation}
where $2M^2/\pi$ is from the phase-space reduction, and
$E_{\gamma}=(m_{\psi(3686)}^2-M^2)/2m_{\psi(3686)}$ is the energy of
the M1 photon in the rest frame of $\psi(3686)$. The empirical damping
function $\mathcal{F}(E_{\gamma})$ is adopted from the study of the KEDR
experiment~\cite{Anashin:2010dh} and given by
\begin{equation}
\mathcal{F}(E_{\gamma})=\frac{E_{\rm on}^2}{E_{\gamma}E_{\rm on}+(E_{\gamma}-E_{\rm on})^2},\label{BkgSlct_eq_chicjDamp}
\end{equation}
where $E_{\rm on}$ is the energy of the photon calculated at the
physical mass of $\eta_{c}(2S)$.

The Gaussian resolution functions $F_{\rm res}^{(a)}$ and
$F_{\rm res}^{(b)}$ in Eq.~\ref{eq6} describe the resolution from
simulation and the difference of resolution between the MC sample and
data, respectively. The former is obtained from MC simulation, while
the latter is fixed by linearly extrapolating $\delta m_2$ and
$\sigma_2$ from the corresponding values of $\chi_{cJ}$. The
$\Gamma^{\text{decay}}$ term in Eq.~\ref{eq7} represents the
energy-dependent decay width of $\eta_c(2S)\to p\bar{p}K^+K^-$. Its
lineshape is assumed to follow the phase space distribution. A signal MC
sample of $\psi(3686)\to\gamma p\bar{p}K^+K^-$ with 0.5 million events
is generated by the phase-space model. Then the GPR method is used to
infer the actual $\Gamma^{\text{decay}}$ function from the
discrete mass distribution of the MC truth. The differential width
${\rm d}\Gamma/{\rm d}M$ after the GPR is related to $\Gamma^{\text{decay}}$ by
\begin{equation}
    \frac{{\rm d}\Gamma}{{\rm d}M}\sim\frac{2M^2}{\pi}E_{\gamma}(M)\Gamma^{\text{decay}}.
\label{fitmethod_eq_edwidth}
\end{equation}
The freedom of choosing $\Gamma^{\text{decay}}$ seems to be unavoidable before we know branching fractions of all intermediate decay modes, which is only achievable with much larger statistics but not possible in this study.

The invariant mass distributions from the exclusive MC samples of
$\psi(3686)\to\gamma\chi_{cJ}\to\gamma p\bar{p}K^+K^-$ are used to
model the shapes of the $\chi_{cJ}$ signal. The decays of $\chi_{cJ}$ are
assumed to have uniform probability throughout the phase space.
Furthermore, to account for the energy dependence in the E1-transition
vertex, the PDFs of the $\chi_{cJ}$ signals are weighted event by event using the
function $w$, given by
\begin{equation}
w=\left(\frac{E_{\gamma}(M_t)}{E_{\rm on}}\right)^3\times \mathcal{F}(E_{\gamma}(M_t)),
\end{equation}
where $M_t$ is the invariant mass of $p\bar{p}K^+K^-$ from the MC
truth. The form of damping function is the same as
Eq.~(\ref{BkgSlct_eq_chicjDamp}), except that $E_{\gamma}$ is
calculated from the physical mass and $E_{\rm on}$ is replaced by the energy
of the photon recoiling against on-shell $\chi_{cJ}$. These PDFs are
convolved with Gaussian functions to account for the resolution
differences between the data and MC simulation. Due to the low background
of this mode, the mass shift and resolution in the single Gaussians
can be obtained by fitting the $M_{p\bar{p}K^+K^-}^{\rm 3C}$
distribution of data around the peaks of $\chi_{cJ}$, and they are
fixed in the final fit using full PDF.

The PDF of $\psi(3686)\to p\bar{p}K^+K^-$ is constructed using the
exclusive MC sample of the $\psi(3686)\to p\bar{p}K^+K^-$ process,
corrected by $R_{\rm FSR}$, and convolved with a single Gaussian with
floating mass shift and resolution.

In the default fit, the mass and width of the $\eta_c(2S)$ are fixed
to their known values~\cite{ParticleDataGroup:2024cfk}. As a cross check, we also perform a fit with
floating masses and widths and obtain values that are
consistent with the known mass and width of $\eta_c(2S)$ within statistical
uncertainties. An input-output check is performed, confirming the
consistency between the output yields of $\eta_c(2S)$ and $\chi_{cJ}$
and the corresponding inputs.

\begin{figure*}[htbp]
 \centering
 \includegraphics[width=0.49\textwidth]{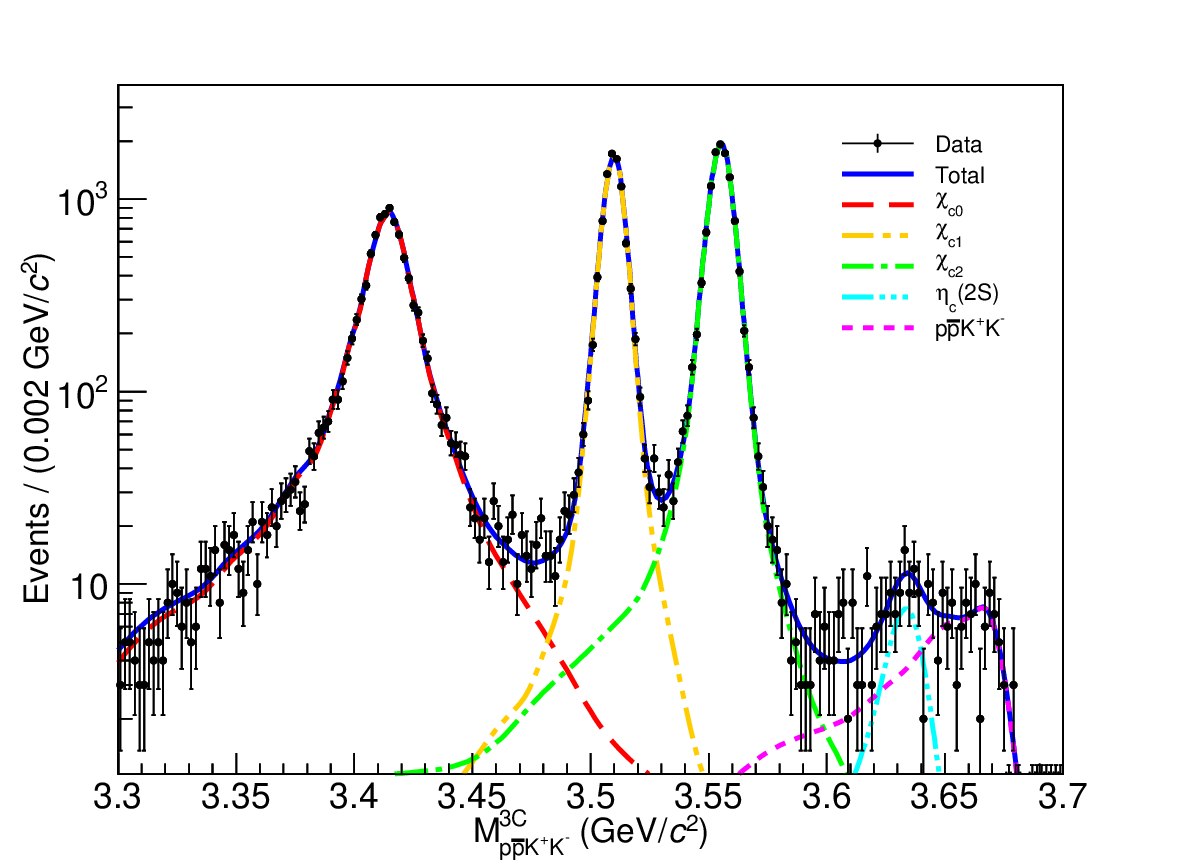}
 \includegraphics[width=0.49\textwidth]{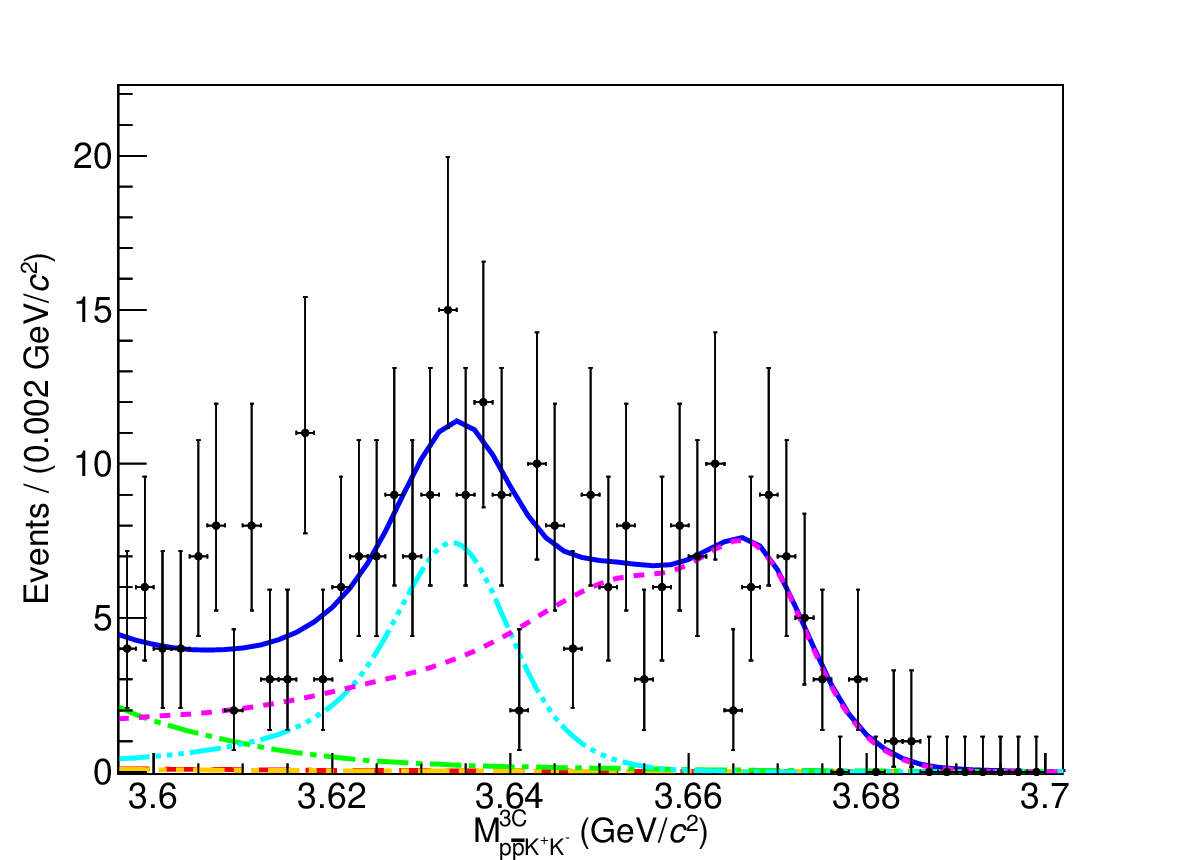}
\caption{The $M_{p\bar{p}K^+K^-}^{\rm 3C}$ distribution after the 3C kinematic fit. The points with error bars are data, and the blue solid, red long-dashed, yellow dash-dot-dotted, green dash-dotted and cyan dash-dot-dot-dotted lines represent the full PDF, PDFs of $\chi_{c0}$, $\chi_{c1}$, $\chi_{c2}$ and $\eta_{c}(2S)$, respectively. The purple short-dashed line is from the $\psi(3686)\to p\bar{p}K^+K^-$ background. The fit result in the signal region of $\eta_c(2S)$
    is displayed in the right panel. }
\label{fitmethod_plot_fitresult}
\end{figure*}

The $M_{p\bar{p}K^+K^-}^{3C}$ distribution and the fit is shown in
Fig.~\ref{fitmethod_plot_fitresult}. Table~\ref{tab:fitresults} shows
the signal yields of the $\eta_c(2S)$ and $\chi_{cJ}$ decays. The statistical
significance of $\eta_c(2S)$ is $3.7\sigma$, calculated by
$\sqrt{-2\ln{(\mathcal{L}/\mathcal{L}_{\rm max})}}$, where
$\mathcal{L}_{\rm max}$ and $\mathcal{L}$ are the likelihoods with and
without the signal of $\eta_c(2S)$, respectively. The product
branching fractions shown in the last column of
Table~\ref{tab:fitresults} are calculated using
\begin{equation}
\mathcal{B}[\psi(3686)\to\gamma X\to\gamma p\bar{p}K^+K^-] =\frac{N_X}{N^{\rm tot}_{\psi(3686)}\epsilon_X},
\end{equation}
where $N_X$ is the fitted yields of state $X\in\{\chi_{cJ}, \eta_c(2S)\}$, $\epsilon_X$ is the detection efficiency and $N^{\rm tot}_{\psi(3686)}=(2712.4\pm14.3)\times 10^6$ is the number of $\psi(3686)$ events in data~\cite{BESIII:2024lks}.

\begin{table*}[htp]
\renewcommand\arraystretch{1.2}
			\caption{\label{tab:fitresults}The measured branching fractions of $\psi(3686)\to\gamma \eta_c(2S)\to\gamma p\bar{p}K^+K^-$ and $\psi(3686)\to\gamma\chi_{cJ}\to\gamma p\bar{p}K^+K^-$. In the second column are signal yields with statistical uncertainties.}
			\centering
			\begin{tabular}{c|c|c}
			\hline\hline
			Mode & $N_X$ & Branching fraction\\
			\hline
			 $\psi(3686)\to\gamma\eta_{c}(2S)\to\gamma p\bar{p}K^+K^-$ & $84\pm 17$ & $(1.98\pm 0.41\pm 0.99)\times 10^{-7}$ \\
			 $\psi(3686)\to\gamma\chi_{c0}\to\gamma p\bar{p}K^+K^-$ & $9952\pm101$ &$(2.49\pm0.03\pm 0.15)\times 10^{-5}$\\
			 $\psi(3686)\to\gamma\chi_{c1}\to\gamma p\bar{p}K^+K^-$ & $8721\pm95$  &$(1.83\pm0.02\pm 0.11)\times 10^{-5}$ \\
			 $\psi(3686)\to\gamma\chi_{c2}\to\gamma p\bar{p}K^+K^-$ & $11463\pm108$  &$(2.43\pm0.02\pm0.15)\times 10^{-5}$ \\
 			\hline\hline
			\end{tabular}
\end{table*}

\section{Systematic uncertainties}
The systematic uncertainties are categorized into two types, i.e. the
additive and the multiplicative uncertainties. These uncertainties are
listed in Table~\ref{tab:sysErr_tab_all}, and the details are
discussed below.

\subsection{The multiplicative uncertainties}
The uncertainties of the tracking and PID of proton and kaon are
studied in Refs.~\cite{BESIII:2019hdp,BESIII:2023lcc} using
$e^+e^-\to p\bar{p}\pi^+\pi^-$ and $e^+e^-\to \pi^+\pi^-K^+K^-$ as
control samples, and the relative uncertainties of tracking and PID for both
particles are estimated to be $1\%$ per track.

The uncertainty associated with the reconstruction of photons is
investigated using control samples of $J/\psi\to \rho^0\pi^0$ and
$e^+e^-\to 2\gamma$~\cite{BESIII:2010ank}, from which the uncertainty
in photon reconstruction is estimated to be $1\%$ per photon. Another
study of $e^+e^-\to\gamma\mu^+\mu^-$~\cite{Prasad:2015bra} process
using $J/\psi$ and $\psi(3770)$ data taken during 2009-2012 also shows
this systematic uncertainty to be $1\%$ at most. Applying the same
method to 9 billion $J/\psi$ events collected in 2009, 2018 and 2019,
the uncertainty in the photon reconstruction efficiency is evaluated
to be $0.5\%$ for photons with energy in the range of [0.1,0.2]
GeV. Since this corresponds to the energy range of photons recoiling
against the $\chi_{cJ}$ states, the photon reconstruction uncertainty
of $0.5\%$ is adopted for the $\chi_{cJ}$ states and $1\%$ for the photons recoiling
against $\eta_c(2S)$, which have lower energy.

The uncertainty from the sampling of fake photons is estimated through
varying the ratio $r$ in each bin by $\pm1\sigma$ around its nominal
value, and the maximum difference in the efficiency is taken as the
systematic uncertainty.

In the kinematic fit, the corrected helix parameters are used in the
nominal result. To quantify the uncertainty associated with the helix
parameter correction, half of the difference in efficiency with and
without the helix parameter correction is taken as the systematic
uncertainty.

The uncertainty in the total number of $\psi(3686)$ events, determined
with inclusive hadronic $\psi(3686)$ decays, is
$0.5\%$~\cite{BESIII:2024lks}. The uncertainty from the quoted branching fractions results from $\mathcal{B}[\chi_{cJ}\to I\to p\bar{p}K^+K^-]$ for different intermediate states $I$.

The value of $\alpha$ in the angular distribution of the photon in $\psi(3680) \to \gamma \chi_{cJ}$ depends on the transition dynamics. Although these processes are dominated by E1 transitions, a small fraction of the amplitude comes from M2 and E3 for $\chi_{c1}$ and $\chi_{c2}$, as measured in Ref.~\cite{BESIII:2017tsq}. MC samples are generated accordingly using the measured values from Ref.~\cite{BESIII:2017tsq}. The difference in efficiency is taken as the systematic uncertainty..

\subsection{The additive uncertainties}
The primary background for the $\eta_c(2S)$ signal arises from the
FSR process, which is sensitive to the value of the
FSR correction factor. To assess the uncertainty from the lineshape of
the background, the FSR factor is varied by $\pm1\sigma$. The largest
deviation in the fitted signal yield is then considered as the
systematic uncertainty.

To estimate the uncertainty associated with non-resonant
contributions, a first-order polynomial is added in the fit
procedure. The resulting fit shows negligible polynomial background. The
change in the log likelihood after the inclusion of the polynomial
contribution is less than 0.001. Therefore, the polynomial background
is ignored in the nominal result, which improves the stability of the
fit. The difference in the signal yields with and without the
polynomial is taken as the systematic uncertainty.

The damping function is changed to
$\exp(-E_{\gamma}^2/(8\beta^2))$, which was used by
CLEO~\cite{CLEO:2008pln} assuming this form factor to depict the
dynamics of the production vertex from $\psi(3686)$. The value of
$\beta$ is fixed at $\beta = 0.12$ GeV, which is obtained from the fit
result of $\chi_{cJ}\to2(\pi^+\pi^-)$ and
$\eta_c(2S)\to 2(\pi^+\pi^-)$ in $\psi(3686)\to\gamma\chi_{cJ}$ and
$\psi(3686)\to\gamma\eta_c(2S)$.  The difference between the fitted
signal yields are taken as the systematic uncertainty due to the damping
function.

In this analysis, the fitting range is from 3.3 GeV to 3.7 GeV. The
Barlow test~\cite{Barlow:2002yb} is performed to test the consistency
of fitted signal yields in different fit ranges. The relative
uncertainty for the branching fraction of the $\chi_{c0}$ signal is $0.12\%$,
while no systematic uncertainty is assigned for other states.

To estimate the systematic uncertainty arising from the efficiency
curve, one hundred samples of the efficiency curves are generated
using the covariance and mean function of the GPR model. The samples
of discrete points are interpolated linearly to replace the nominal
efficiency curves. The distribution of events obtained using these
alternative interpolated curves are fitted with a normal distribution,
whose standard deviation is taken as the systematic uncertainty.

The difference between the MC simulation and the data is characterized
by the mass shift $\delta m_2$ and the resolution $\sigma_2$. The
$\delta m_2$($\sigma_2$) of $\eta_c(2S)$ is estimated by fitting the
$\delta m_2$($\sigma_2$) of $\chi_{cJ}$ with a first-order polynomial
and extrapolate to the mass of $\eta_c(2S)$. The uncertainty of
$\sigma_2$ arising from this linear assumption is estimated by
comparing the nominal result with the one with increasing 
$\sigma_2$ by one standard deviation. To estimate the uncertainty of
$\delta m_2$, its value is varied by one standard deviation in
both directions, and the larger difference in the fitted signal
yields of $\eta_c(2S)$ is taken as the systematic uncertainty. Since
 $\delta m_2$ and $\sigma_2$ of $\eta_c(2S)$ are strongly
positively correlated, the total systematic uncertainty is obtained by
adding numbers from these two sources linearly.

\begin{table*}[htp]
\renewcommand\arraystretch{1.4}
\centering
\caption{\label{tab:sysErr_tab_all}  Systematic uncertainties in product branching fractions. The total uncertainty is taken as the sum of all terms in quadrature.}
\begin{tabular}{l|c|c|c|c}
\hline\hline
Item & $\eta_c(2S)$ & $\chi_{c0}$ & $\chi_{c1}$ & $\chi_{c2}$ \\
\hline
Tracking & $4.0\%$ & $4.0\%$ & $4.0\%$ & $4.0\%$\\
PID & $4.0\%$ & $4.0\%$ & $4.0\%$ & $4.0\%$\\
Photon reconstruction & $1.0\%$ & $0.5\%$ & $0.5\%$ & $0.5\%$\\
Photon sampling & $0.8\%$ & $0.2\%$ & $1.3\%$ & $1.3\%$\\
Kinematic fit & $2.1\%$ & $1.6\%$ & $1.8\%$ & $1.9\%$\\
$N_{\psi(3686)}$ & $0.5\%$ & $0.5\%$ & $0.5\%$ & $0.5\%$\\
$\mathcal{B}$ from PDG & - & $0.4\%$ & $0.5\%$ & $0.5\%$ \\
$\alpha$  & - & - & $0.1\%$ & $0.9\%$ \\
\hline
FSR & $16\%$ & - & - & $0.1\%$\\
Background shape & $0.2\%$ & - & - & -\\
Damping function & $47\%$ & $0.1\%$ & $0.1\%$ & $0.1\%$\\
Fit range & - & $0.1\%$ & - & -\\
Efficiency curve & $0.5\%$ & - & - & -\\
Resolution & $1.4\%$ & - & - & -\\
\hline
Total & $50\%$ & $6.0\%$ & $6.2\%$ & $6.2\%$ \\
\hline\hline
\end{tabular}
\end{table*}

\section{Results and discussion}
The measured branching fractions are shown in
Table~\ref{tab:fitresults}. The significance of the $\eta_c(2S)$
signal is $3.3\sigma$ after considering the systematic uncertainty.
To obtain a conservative estimation for the upper limit of
$\mathcal{B}[\psi(3686)\to\gamma\eta_c(2S)\to\gamma p\bar{p}K^+K^-]$,
the FSR factor $R_{\rm FSR}$ is set to a value one standard deviation lower than
the nominal one, and the CLEO form factor with $\beta=0.12$ GeV is
used. The curve of likelihood, as a function of $N$, is then convolved
with a single Gaussian via the following formula:
\begin{equation}
\tilde{L}(N)=\int_0^1L\left(\frac{\epsilon}{\hat{\epsilon}}N\right)\frac{1}{\sqrt{2\pi\sigma^2_{\epsilon}}}e^{-\frac{(\epsilon-\hat{\epsilon})^2}{2\sigma_{\epsilon}^2}}d\epsilon\ .
\end{equation}
Here, $\hat{\epsilon}$ represents the nominal efficiency, and
$\sigma_{\epsilon}$ denotes the multiplicative systematic uncertainty
associated with it.  The upper limit of
$\mathcal{B}[\psi(3686)\to\gamma\eta_c(2S)\to\gamma p\bar{p}K^+K^-] $
at  $90\%$ C.L. is
set to be 
$4.1 \times 10^{-7}$.

\begin{table*}[htbp]
\renewcommand\arraystretch{1.4}
\centering
\caption{The branching fractions of $\chi_{cJ}\to p\bar{p}K^+K^-$ through different modes measured before this work~\cite{ParticleDataGroup:2024cfk}.}
\begin{tabular}{c|c|c|c|c}
\hline\hline
State & non-resonant & $\Lambda(1520)\bar{\Lambda}(1520)$ & $K^+\bar{p}\Lambda(1520)+\text{c.c.}$ & $p\bar{p}\phi$ \\
\hline
$\chi_{c0}$ & $(1.2\pm0.3)\times 10^{-4}$  & $(3.1\pm1.2)\times 10^{-4}$ & $(3.0\pm0.8)\times 10^{-4}$ & $(6.0\pm1.4)\times 10^{-5}$\\
\hline
$\chi_{c1}$ & $(1.3\pm0.2)\times 10^{-4}$  & $<9\times 10^{-5}$ & $(1.7\pm0.4)\times 10^{-4}$ & $<1.7\times 10^{-5}$\\
\hline
$\chi_{c2}$ & $(1.9\pm0.3)\times 10^{-4}$  & $(4.7\pm1.5)\times 10^{-4}$ & $(2.9\pm0.7)\times 10^{-4}$ & $(2.8\pm0.9)\times 10^{-5}$\\
\hline\hline
\end{tabular}
\label{fitmethod_tab_br_imed}
\end{table*}

Since the product branching fractions of
$\psi(3686)\to\gamma\chi_{cJ}\to\gamma p\bar{p}K^+K^-$ were not
reported in the previous study using 106 million $\psi(3686)$
events~\cite{Wang:2011kti}, the branching fractions of
$\mathcal{B}[\chi_{cJ}\to p\bar{p}K^+K^-]$ are estimated by
incoherently summing the measured branching fractions shown in
Table~\ref{fitmethod_tab_br_imed}. To estimate the uncertainty of the
incoherent sum, the branching fractions in
Table~\ref{fitmethod_tab_br_imed} are assumed to be independent on
each other. In the calculation,
$\mathcal{B}[\Lambda(1520)\to pK^-]=0.23\pm (0.01\times 0.23/0.45)$ is
assumed, where the 0.45 is the branching fraction of
$\Lambda(1520)\to N\bar{K}$ from the PDG~\cite{ParticleDataGroup:2024cfk}. For
$\phi\to K^+K^-$, $\mathcal{B}[\phi\to K^+K^-]=0.491\pm 0.001$ is
adopted. For values with upper limit only, the measured value is
treated to be zero and the upper limit is taken as the
uncertainty. The results are shown in Table~\ref{sysErr_tab_br_comp},
where the branching fractions calculated from this work together with
the branching fractions of $\psi(3686)\to\gamma\chi_{cJ}$ are also
listed. All branching fractions are assumed to be independent on each
other. The branching fractions from this work agree with
those in Ref.~\cite{Wang:2011kti}, while the precision is improved. Given the
current large statistics, a partial wave analysis is necessary to
obtain more reliable measurements of the branching fractions of these
decay modes of $\chi_{cJ}$, which is beyond the scope of this
paper. Nevertheless, the previous measurement can be considered as a
rough estimation and is complementary to the measurement of product
branching fractions in this work.

\begin{table*}
\renewcommand\arraystretch{1.4}
\centering
\caption{The branching fractions of $\chi_{cJ}\to p\bar{p}K^+K^-$ from Ref.~\cite{Wang:2011kti} and this work. The branching fractions of $\psi(3686)\to\gamma\chi_{cJ}$ are quoted from PDG~\cite{ParticleDataGroup:2024cfk}. The uncertainties are combined.}
\begin{tabular}{c|c|c|c}
\hline\hline
State & $\mathcal{B}[\psi(3686)\to\gamma\chi_{cJ}]$ & Ref.~\cite{Wang:2011kti} & This work \\
\hline
$\chi_{c0}$ & $(9.77\pm0.23)\%$  & $(2.3\pm 0.3)\times 10^{-4}$ & $(2.55\pm 0.17)\times 10^{-4}$ \\
\hline
$\chi_{c1}$ & $(9.75\pm0.27)\%$  & $(1.7\pm 0.3)\times 10^{-4}$ & $(1.87\pm 0.13)\times 10^{-4}$ \\
\hline
$\chi_{c2}$ & $(9.36\pm0.23)\%$  & $(2.9\pm 0.6)\times 10^{-4}$ & $(2.59\pm 0.17)\times 10^{-4}$ \\
\hline\hline
\end{tabular}
\label{sysErr_tab_br_comp}
\end{table*}

\section{summary}
In summary, based on $(2712.4\pm14.3)\times 10^6$ $ \psi(3686)$ events
collected with the BES\uppercase\expandafter{\romannumeral3} detector
at BEPC\uppercase\expandafter{\romannumeral2}, the branching fractions
of
$\psi(3686)\to\gamma \eta_c(2S), \gamma\chi_{cJ}\to \gamma
p\bar{p}K^+K^-$ are measured. The first evidence for the $\eta_c(2S)$
in this mode is found with a significance of $3.3\sigma$ after
considering the systematic uncertainties. The branching fraction of
$\psi(3686)\to\gamma\eta_c(2S)\to \gamma p\bar{p}K^+K^-$ is measured
to be
$(1.98\mkern 2mu\pm \mkern 2mu 0.41\mkern 2mu \pm \mkern 2mu
0.99)\times 10^{-7}$, and its upper limit at $90\%$ C.L. is set to be
$4.1\times10^{-7}$.  The branching fractions of
$\psi(3686)\to\gamma\chi_{cJ}\to \gamma p\bar{p}K^+K^-$ are determined
to be
$(2.49\mkern 2mu\pm\mkern 2mu 0.03\mkern 2mu\pm\mkern 2mu0.15)\times
10^{-5}$,
$(1.83\mkern 4mu\pm\mkern 4mu 0.02\mkern 4mu\pm\mkern 4mu0.11)\times
10^{-5}$, and
$(2.43\mkern 2mu\pm\mkern 2mu 0.02\mkern 2mu\pm\mkern 2mu0.15)\times
10^{-5}$, for $\chi_{c0}$, $\chi_{c1}$, and $\chi_{c2}$,
respectively. Benefiting from the high statistics of $\chi_{cJ}$ and
negligible background, these decay modes hold great promise to achieve a
deeper understanding of their decay dynamics, especially the baryons
and light mesons interactions, in further studies.

\acknowledgments

The BESIII Collaboration thanks the staff of BEPCII and the IHEP computing center for their strong support. This work is supported in part by National Key R\&D Program of China under Contracts Nos. 2020YFA0406300, 2020YFA0406400, 2023YFA1606000; National Natural Science Foundation of China (NSFC) under Contracts Nos. 11635010, 11735014, 11935015, 11935016, 11935018, 12025502, 12035009, 12035013, 12061131003, 12192260, 12192261, 12192262, 12192263, 12192264, 12192265, 12221005, 12225509, 12235017, 12361141819; the Chinese Academy of Sciences (CAS) Large-Scale Scientific Facility Program; the CAS Center for Excellence in Particle Physics (CCEPP); Joint Large-Scale Scientific Facility Funds of the NSFC and CAS under Contract No. U1832207; 100 Talents Program of CAS; The Institute of Nuclear and Particle Physics (INPAC) and Shanghai Key Laboratory for Particle Physics and Cosmology; German Research Foundation DFG under Contracts Nos. FOR5327, GRK 2149; Istituto Nazionale di Fisica Nucleare, Italy; Knut and Alice Wallenberg Foundation under Contracts Nos. 2021.0174, 2021.0299; Ministry of Development of Turkey under Contract No. DPT2006K-120470; National Research Foundation of Korea under Contract No. NRF-2022R1A2C1092335; National Science and Technology fund of Mongolia; National Science Research and Innovation Fund (NSRF) via the Program Management Unit for Human Resources \& Institutional Development, Research and Innovation of Thailand under Contracts Nos. B16F640076, B50G670107; Polish National Science Centre under Contract No. 2019/35/O/ST2/02907; Swedish Research Council under Contract No. 2019.04595; The Swedish Foundation for International Cooperation in Research and Higher Education under Contract No. CH2018-7756; U. S. Department of Energy under Contract No. DE-FG02-05ER41374

\clearpage{}\clearpage{}

\end{document}